\documentclass[epj]{webofc}
\usepackage[utf8]{inputenc}
\usepackage[varg]{txfonts}   
\usepackage{booktabs}
\usepackage{xcolor}
\definecolor{darkred}{rgb}{0.4,0.0,0.0}
\definecolor{darkgreen}{rgb}{0.0,0.4,0.0}
\definecolor{darkblue}{rgb}{0.0,0.0,0.4}
\usepackage[bookmarks,linktocpage,colorlinks,
    linkcolor = darkred,
    urlcolor  = darkblue,
    citecolor = darkgreen]{hyperref}
\wocname{EPJ Web of Conferences}
\woctitle{Lattice2017}
\usepackage{braket}
\usepackage{multirow}
\usepackage{dsfont}
\usepackage{mathtools}
\usepackage{xcolor}
\usepackage[normalem]{ulem}
\begin{document}
%
\selectlanguage{english}
\title{%
Degeneracy of vector-channel spatial correlators \\ in high temperature QCD 
}
\author{%
\firstname{Christian} \lastname{Rohrhofer}\inst{1}\fnsep\thanks{Speaker, \email{christian.rohrhofer@uni-graz.at}} \and
\firstname{Yasumichi} \lastname{Aoki}\inst{2,3} \and
\firstname{Guido}  \lastname{Cossu}\inst{4} \and
\firstname{Hidenori} \lastname{Fukaya}\inst{5} \and
\firstname{Leonid} \lastname{Glozman}\inst{1} \and
\firstname{Shoji} \lastname{Hashimoto}\inst{2,6} \and
\firstname{Christian B.} \lastname{Lang}\inst{1} \and
\firstname{Sasa} \lastname{Prelovsek}\inst{7,8,9}
}
\institute{%
Institute of Physics, University of Graz, 8010 Graz, Austria
\and
KEK Theory Center, High Energy Accelerator Research Organization (KEK), Tsukuba 305-0801, Japan
\and
RIKEN BNL Research Center, Brookhaven National Laboratory, Upton NY 11973, USA
\and
School of Physics and Astronomy, The University of Edinburgh, Edinburgh EH9 3JZ, United Kingdom
\and
Department of Physics, Osaka University, Toyonaka 560-0043, Japan
\and
School of High Energy Accelerator Science, The Graduate University for Advanced Studies (Sokendai), Tsukuba 305-0801, Japan
\and
Faculty of Mathematics and Physics, University of Ljubljana, 1000 Ljubljana, Slovenia
\and
Jozef Stefan Institute, 1000 Ljubljana, Slovenia
\and
Institute f\"ur Theoretische Physik, Universit\"at Regensburg, D-93040, Germany
}
\abstract{%
  We study spatial isovector meson correlators in $N_f=2$ QCD with dynamical
  domain-wall fermions on $32^3\times 8$ lattices at temperatures
  up to 380 MeV with various quark masses. We measure the correlators of
  spin-one isovector operators including vector, axial-vector,
  tensor and axial-tensor.
  At temperatures above $T_c$ we observe an approximate degeneracy of the
  correlators in these channels, which is unexpected because some of them are
  not related under $\mathrm{SU}(2)_L \times \mathrm{SU}(2)_R$ nor $\mathrm{U}(1)_A$ symmetries.
  The observed approximate degeneracy suggests emergent $\mathrm{SU}(2)_{CS}$
  (chiral-spin) and $\mathrm{SU}(4)$ symmetries at high $T$.
}
\maketitle
\section{Introduction}\label{intro}

The behaviour of strongly interacting matter at high temperatures and its
description in the phase diagram of QCD are important questions for a better
understanding of the theory as well as for any kind of future applications.
It is a priori not clear how two main features of QCD, confinement and chiral
symmetry breaking, will behave in regions of higher temperature and/or density.
For the latter it has been established that above a critial temperature
QCD becomes chirally symmetric.
In this regime the effects of the anomalous breaking of the $\mathrm{U}(1)_A$
symmetry are suppressed.
There is strong evidence from lattice calculations using chirally-invariant
fermions that the $\mathrm{U}(1)_A$ symmetry gets restored along with a
gap opening in the eigenmode spectrum of the Dirac operator
\cite{Cossu:2013uua,Tomiya:2016jwr,Bazavov:2012qja}.

In this work we calculate spatial correlation functions of quark bilinears
to probe their symmetry structure. The operators of interest are all possible
local isovectors with spin $J=0$ and $J=1$. As fermion discretization we
use the chirally invariant domain-wall formalism, and calculate the correlations
at four temperatures from $\sim 1.2 T_c$ to $\sim 2.2 T_c$.
The motivation for this is two-fold:
\begin{itemize}
\item
In the high temperature, chirally symmetric regime correlation functions
connected by $\mathrm{SU}(2)_L \times \mathrm{SU}(2)_R$ transformations are
expected to become degenerate. The effects of $\mathrm{U}(1)_A$ breaking are expected
to be suppressed. This regime eventually becomes $\mathrm{U}(2)_L \times \mathrm{U}(2)_R$
symmetric \cite{Cohen:1996ng}.
\item
In a series of numerical studies at $T=0$, where near-zero modes of the Dirac operator
have been truncated from the quark propagators, extended
$\mathrm{SU}(2)_{CS}$ (chiral-spin) and $\mathrm{SU}(4)$ symmetries
\cite{Glozman:2014mka, Glozman:2015qva}
have been observed in the
hadron spectrum
\cite{Denissenya:2014poa,Denissenya:2014ywa,Denissenya:2015mqa,Denissenya:2015woa}.
The low eigenmodes of the Dirac operator are strongly suppressed at
high temperatures \cite{Cossu:2013uua,Tomiya:2016jwr}, which raises the question
whether these symmetries also occur naturally in the high $T$ spectrum. 
\end{itemize}

\section{\label{sec:sim}Lattices and operators}

The configurations are generated using the
Symanzik action for the gauge sector and the 
M\"obius domain wall formalism as fermion discretization
\cite{Brower:2005qw,Brower:2012vk}. The length of the fifth dimension for
the domain wall fermions is chosen separately for each ensemble in a manner
which guarantees good chiral properties at moderate cost.
The gauge links are stout smeared three times before the computation of the
Dirac operator; the boundary conditions for quarks are set antiperiodic in
$t-$direction and periodic in spatial directions.
The ensembles and parameters including the lattice spacing $a$ are listed
in Table~\ref{tab:ensembles}, see also \cite{Cossu:2015kfa,Tomiya:2016jwr}.

\begin{table}
\center
\begin{tabular}{c|c|c|c|c|c|c}
\hline \hline
$\beta$ & $m_{ud}a$ & $a$ [fm] & \# configs & $L_s$ & $T$ [MeV] & $T/T_c$ \\\hline
$4.10$  & 0.001 & $0.113$ &     800  & 24  & $\sim220$ & $\sim 1.2$ \\
$4.18$  & 0.001 & $0.096$ &     230  & 12  & $\sim260$ & $\sim 1.5$ \\
$4.30$  & 0.001 & $0.075$ &     260  & 12  & $\sim320$ & $\sim 1.8$ \\
$4.37$  & 0.005 & $0.065$ &     120  & 12  & $\sim380$ & $\sim 2.2$ \\
\hline \hline
\end{tabular}
\caption{Ensembles for $32^3 \times 8$ lattices used in this work.
$L_s$ is the length of the fifth dimension in the domain wall fermion
formulation. The critical temperature for this set of parameters
is $T_c=175\pm5$~MeV. The mass range of degenerate up and down quarks
$m_{ud}$ is 2--15~MeV.}
\label{tab:ensembles}
\end{table}


As observables we use correlators of local isovector operators
$\mathcal{O}_\Gamma(x) = \bar q(x) \Gamma \frac{\vec{\tau}}{2} q(x)$,
where $\Gamma$ might be any element of the Clifford algebra;
$\tau_a$ are the isospin Pauli matrices.
We study spatial correlators in $z$-direction, as suggested in ref.
\cite{DeTar:1987xb}.
Therefore we use a zero-momentum
projection by summing over all lattice points in slices orthogonal
to the measurement direction:
\begin{equation}
C_\Gamma(n_z) = \sum\limits_{n_x, n_y, n_t}
\braket{\mathcal{O}_\Gamma(n_x,n_y,n_z,n_t)
\mathcal{O}_\Gamma(\mathbf{0},0)^\dagger}.
\label{eq:momentumprojection}
\end{equation}
Thus we can identify the propagating components of the Clifford algebra and
organize them to spin $J=0$ and $J=1$ operators:
The Pseudoscalar ($PS$) is given by $\Gamma=\gamma_5$, and the Scalar ($S$) operator by
$\Gamma=\mathds{1}$.
For the Vector and Axial-vector operators $\Gamma$ has the
following components:
\begin{align}
\mathbf{V}=
\left(
\begin{array}{c}
\gamma_1 = V_x \\
\gamma_2 = V_y \\
\gamma_4 = V_t 
\end{array}
\right),
\quad
\mathbf{A}=
\left(
\begin{array}{c}
\gamma_1 \gamma_5 = A_x \\
\gamma_2 \gamma_5  = A_y \\
\gamma_4 \gamma_5  = A_t 
\end{array}
\right).
\label{eq:vectorcurrents}
\end{align}
Conservation of the vector current requires that $V_z$ does not propagate in
$z$-direction. 
As the axial vector current $j_5^\mu$ is not conserved at zero temperature,
the relevant component $\gamma_3 \gamma_5$ of the Axial-vector does
propagate at zero temperature and eventually couples to the
Pseudoscalar. Above the critical temperature --- after $\mathrm{U}(1)_A$ and
$\mathrm{SU}(2)_L \times \mathrm{SU}(2)_R$ restoration ---
$A_z$ behaves as its parity partner $V_z$ and does not propagate in
$z$-direction.
For propagation in $z$-direction the tensor elements $\sigma_{\mu\nu}$
of the Clifford algebra are organized
in the following way in components of Tensor- and
Axial-tensor operators:
\begin{align}
\mathbf{T}=
\left(
\begin{array}{c}
\gamma_1 \gamma_3 = T_x \\
\gamma_2 \gamma_3 = T_y \\
\gamma_4 \gamma_3 = T_t 
\end{array}
\right),
\quad
\mathbf{X}=
\left(
\begin{array}{c}
\gamma_1 \gamma_3 \gamma_5 = X_x \\
\gamma_2 \gamma_3 \gamma_5 = X_y \\
\gamma_4 \gamma_3 \gamma_5 = X_t 
\end{array}
\right).
\label{tensorcurrents}
\end{align}
Table~\ref{tab:ops} summarizes our operators
and gives the  $\mathrm{U}(1)_A$ and $\mathrm{SU}(2)_L \times \mathrm{SU}(2)_R$ relations of these operators. Given
restoration of the $\mathrm{U}(1)_A$ and $\mathrm{SU}(2)_L \times \mathrm{SU}(2)_R$ symmetries at high $T$ we expect
degeneracies of correlators calculated with the corresponding operators.

\begin{table}
\center
\begin{tabular}{lccll}
\hline\hline
 Name        &
 Dirac structure &
 Abbreviation    &
 \multicolumn{2}{l}{
 } \\\hline
\textit{Pseudoscalar}        & $\gamma_5$                 & $PS$         & \multirow{2}{*}{$\left.\begin{aligned}\\ \end{aligned}\right\} \mathrm{U}(1)_A$} &\\
\textit{Scalar}              & $\mathds{1}$               & $S$          & &\\\hline
\textit{Axial-vector}        & $\gamma_k\gamma_5$         & $\mathbf{A}$ & \multirow{2}{*}{$\left.\begin{aligned}\\ \end{aligned}\right\} \mathrm{SU}(2)_A$}&\\
\textit{Vector}              & $\gamma_k$                 & $\mathbf{V}$ & & \\
\textit{Tensor-vector}       & $\gamma_k\gamma_3$         & $\mathbf{T}$ & \multirow{2}{*}{$\left.\begin{aligned}\\ \end{aligned}\right\} \mathrm{U}(1)_A$} &\\
\textit{Axial-tensor-vector} & $\gamma_k\gamma_3\gamma_5$ & $\mathbf{X}$ & &\\
\hline\hline
\end{tabular}
\caption{
Bilinear operators considered in this work and their transformation properties
(last column). This classification assumes propagation in $z$-direction. The
open vector index $k$ denotes the components $1,2,4$, \textit{i.e.} $x,y,t$.}
\label{tab:ops}
\end{table}

For measurements at zero temperature the three components of the vectors give
the same expectation value due to the $\mathrm{SO}(3)$ symmetry in continuum.  
On the finite temperature lattices the corresponding symmetry group is $D_{4h}$
where the Vector has one
two-dimensional ($V_x$,$V_y$) and one one-dimensional ($V_t$) irreducible
representation, and similar for $\mathbf{A}$, $\mathbf{T}$, $\mathbf{X}$.
Thus only $x$- and $t$-components are shown subsequently for
the relevant operators.

\section{\label{sec:results}Results}

\begin{figure}
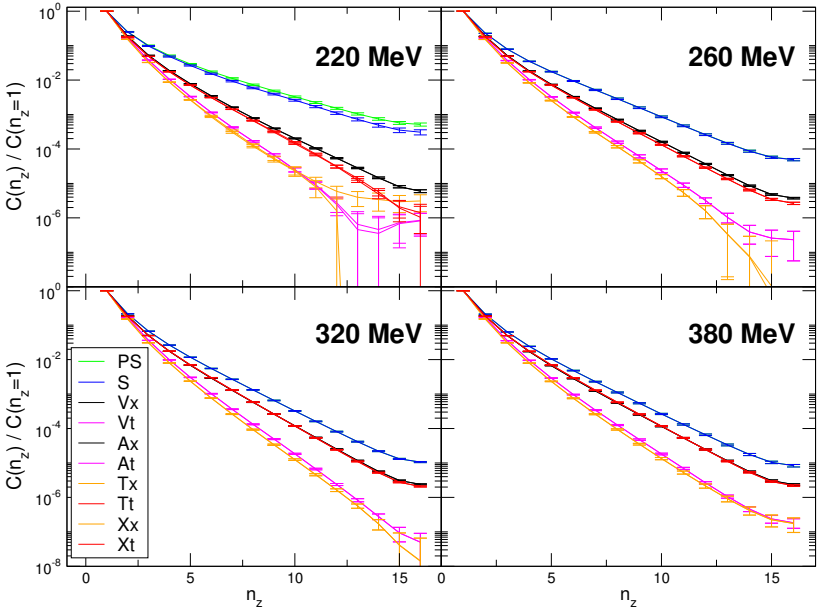

  \centering
  \includegraphics[scale=0.4]{{{1}}}
  \caption{
    Normalized spatial correlators. The temperatures correspond
    to the ensembles listed in Table~\protect{\ref{tab:ensembles}}.
  }
  \label{fig:corrs}
\end{figure}

Figure \ref{fig:corrs} shows the spatial correlation functions normalized to 1
at $n_z=1$ for the operators given in Table~\ref{tab:ops}.
As argument we show $n_z$ which is proportional to the dimensionless product
$zT$ for fixed $N_t$, the temporal extent of the lattice. 

As we describe in more detail later, we find that all correlators 
connected by $\mathrm{SU}(2)_L \times \mathrm{SU}(2)_R$ and $\mathrm{U}(1)_A$ transformations coincide 
within small deviations at $T > 220$~MeV, which means that at these
temperatures both chiral symmetries get restored. 
More interestingly, there are additional degeneracies of correlators.
In total we observe three different multiplets:
\begin{align}
E_1: & \quad PS \leftrightarrow S \label{eq:e1} \\
E_2: & \quad V_x \leftrightarrow T_t \leftrightarrow X_t \leftrightarrow A_x \label{eq:e2} \\
E_3: & \quad V_t \leftrightarrow T_x \leftrightarrow X_x \leftrightarrow A_t. \label{eq:e3}
\end{align}
$E_1$ describes the Pseudoscalar-Scalar multiplet connected
by the $\mathrm{U}(1)_A$ symmetry,
that is realized in the absence of chiral 
zero-modes \cite{Kogut:1998rh}.
Note that we only consider the isospin triplet
channels so $S$ corresponds to the $a_0$- rather than the $\sigma$-particle.
The $E_2$ multiplet on the other hand contains
some operators that are not connected by either
$\mathrm{SU}(2)_L \times \mathrm{SU}(2)_R$ or $\mathrm{U}(1)_A$
transformations; this holds also for the $E_3$ multiplet.

\begin{figure}
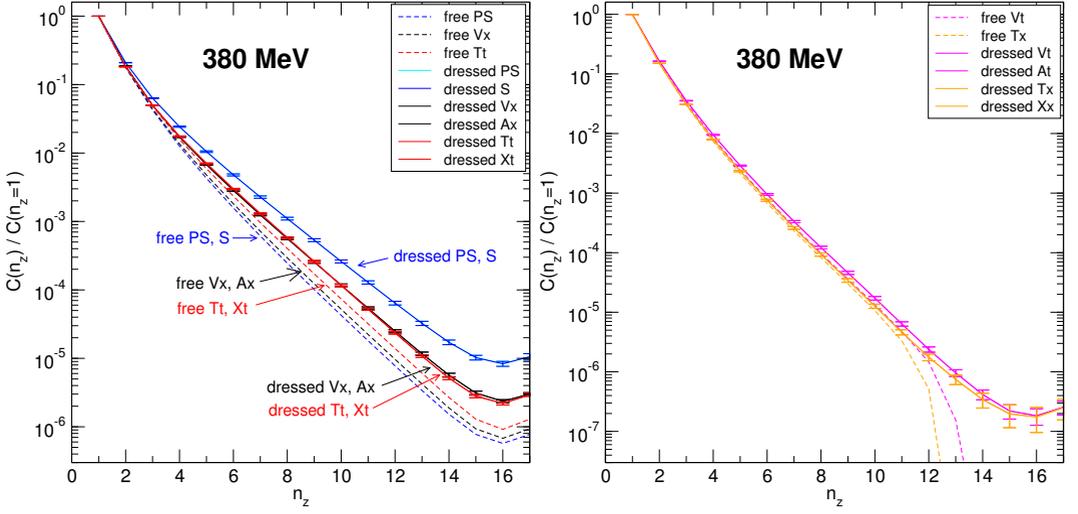

  \centering
  \includegraphics[scale=0.4]{{{2}}}
  \includegraphics[scale=0.4]{{{3}}}
  \caption{
    {\it Left side:}
    $E_1$ and $E_2$ multiplets
    (\protect{\ref{eq:e1}}-\protect{\ref{eq:e2}}) for interacting
    (\textit{dressed}) and non-interacting (\textit{free}) calculations
    at $T$= 380~MeV.
    {\it Right side:}
    $E_3$ multiplet (\protect{\ref{eq:e3}}) for interacting
    (\textit{dressed}) and non-interacting (\textit{free}) calculations.
  }
  \label{fig:e2}
\end{figure}

The left side of Figure \ref{fig:e2} shows the correlators of the $E_1$ and $E_2$ multiplets
in detail at the highest available temperature $T= 380$~MeV.
Here we also show correlators calculated with non-interacting
quarks. The
non-interacting (\textit{free}) data have been generated on the same
lattice sizes
using a unit gauge configuration and verified by analytic calculation.
Due to the small quark mass the differences
between chiral partners is negligible for the free case, therefore they are
omitted.

We observe a precise degeneracy between $S$ and $PS$ correlators, 
which is consistent with the $\mathrm{U}(1)_A$ restoration on these 
lattice ensembles \cite{Cossu:2013uua}.
The logarithmic slope of the interacting (\textit{dressed}) \textit{S}
and \textit{PS} correlators is substantially smaller
than that for free quarks. In the latter case the slope is given
by $2\pi/N_t$. A system of two free quarks cannot have `energy' smaller than
twice the lowest Matsubara frequency \cite{DeTar:1987xb}.
For the $E_2$ multiplet we observe asymptotic slopes that are quite
close to $2\pi/N_t$ in agreement with previous studies \cite{Karsch:2003jg}.

\begin{figure}
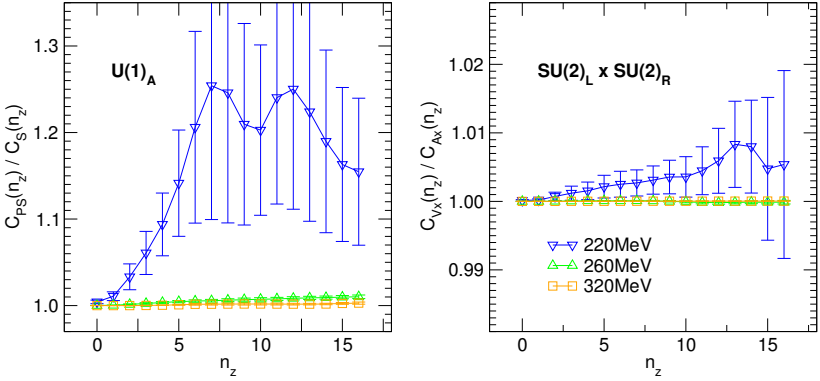

  \centering
  \includegraphics[scale=0.4]{{{comp_cs}}}
  \caption{
    Ratios of $PS$ and $S$ as well as of $V_x$ and $A_x$ correlators from Figure \protect{\ref{fig:corrs}},
    that are related by chiral $\mathrm{U}(1)_A$ and
    $\mathrm{SU}(2)_L\times \mathrm{SU}(2)_R$ transformations.
  }
  \label{fig:ratios1}
\end{figure}

\begin{figure}
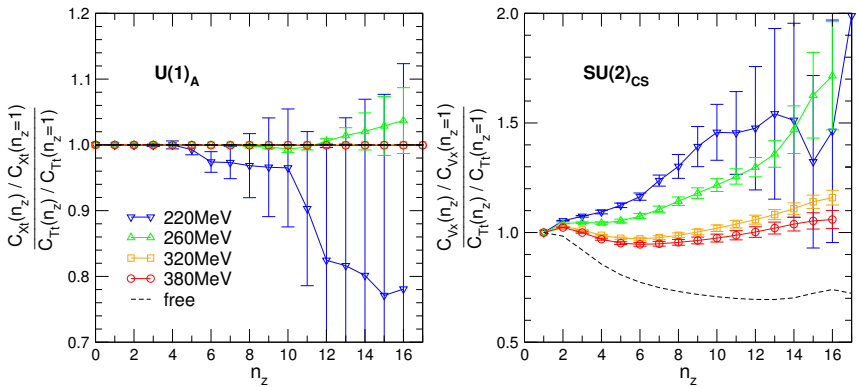

  \centering
  \includegraphics[scale=0.4]{{{4a}}}
  \includegraphics[scale=0.4]{{{4b}}}
  \caption{
    Ratios of normalized $X_t$ and $T_t$ as well as of $V_x$ and $T_t$ correlators from Figure \protect{\ref{fig:corrs}},
    that are related by $\mathrm{U}(1)_A$ and $\mathrm{SU}(2)_{CS}$ transformations.
  }
  \label{fig:ratios2}
\end{figure}

Figures \ref{fig:ratios1} and \ref{fig:ratios2} show
ratios of correlators connected by various symmetries at different temperatures.
We use these ratios as quantities to measure the level of symmetry breaking.
Figure \ref{fig:ratios1} shows ratios of $PS$ and $S$ correlators on the left, and
ratios of $V_x$ and $A_x$ correlators on the right. 
Figure \ref{fig:ratios2} shows ratios of
$X_t$ and $T_t$ correlators on the left,
as well as of $V_x$ and $T_t$ correlators on the right side.
The chiral symmetries $\mathrm{U}(1)_A$ and
$\mathrm{SU}(2)_L \times \mathrm{SU}(2)_R$ are restored at $T>220$ MeV,
as is evident from both Figures
(see also, {\it e.g.} \cite{Cheng:2010fe,Ishikawa:2017nwl}).
Interesting is the level of residual breaking at $T=220$ MeV: The corresponding
deviation from unity
for $U(1)_A$ connected correlators is at least one order of magnitude higher than for
$\mathrm{SU}(2)_L \times \mathrm{SU}(2)_R$ connected correlators.
\footnote{This effect might be spoiled by chirality violating eigenmodes of the Dirac operator,
which are non-neglible for this ensemble and temperature \cite{Tomiya:2016jwr}. They are absent for higher temperatures.} 

Figures \ref{fig:e2} and \ref{fig:ratios2} suggest a possible higher
symmetry that connects $V_x$ and $T_t$ channels. The right 
panel of Figure \ref{fig:ratios2} shows the corresponding ratio, which demonstrates an 
approximate degeneracy at the level of 5\% above $T\simeq$ 320~MeV. We 
notice that this degeneracy is not expected in the free quark limit 
which is plotted by a dashed curve.
This unexpected symmetry requires that the cross-correlator calculated with the
$V_x$ and $T_t$ operators (both create the $1^{--}$ states) should vanish.
We have carefully checked that it indeed vanishes to high accuracy.

The right side of Figure \ref{fig:e2} shows the $E_3$ multiplet. Here again we observe a precise
degeneracy in all $\mathrm{SU}(2)_L \times \mathrm{SU}(2)_R$  and $\mathrm{U}(1)_A$ connected correlators,
as well as the approximate degeneracy in all four correlators.
We also see qualitatively
different data between free and dressed correlators at $n_z \geq 11$,
as also seen in \cite{Gavai:2008yv}. For this set of operators the
spatial correlations of free quarks become negative near the symmetry point,
which is a finite size effect.
The absence of this behaviour in the dressed correlators implies that we
do not observe free non-interacting quarks but instead
systems with some interquark correlation, which is in accordance with
the known results for energy density and pressure at high temperatures
\cite{Karsch:2000ps}.

\section{\label{sec:su4}$\mathrm{SU}(2)_{CS}$ and $\mathrm{SU}(4)$ symmetries}

In this section we introduce the $\mathrm{SU}(2)_{CS}$ and $\mathrm{SU}(4)$ transformations,
which connect operators from multiplet $E_2$ (\ref{eq:e2})
as well as from multiplet $E_3$ (\ref{eq:e3})
and contain chiral transformations as a subgroup.
The basic ideas of $\mathrm{SU}(2)_{CS}$ and $\mathrm{SU}(4)$ symmetries at zero temperature
are given in \cite{Glozman:2015qva}. Here we adapt the group
structure to our setup.

We use the $\gamma$-matrices given by
\begin{align}
\gamma_i\gamma_j + \gamma_j \gamma_i &=
2\delta_{ij} \\
\gamma_5 &= \gamma_1\gamma_2\gamma_3\gamma_4.
\label{eq:diracalgebra}
\end{align}
The generators of the $\mathrm{SU}(2)_{CS}$  chiral-spin group,
defined in the Dirac spinor space and diagonal in flavour space, are given by
\begin{equation}
\vec \Sigma = \{\gamma_k,-i \gamma_5\gamma_k,\gamma_5\}.
\label{eq:su2cs_}
\end{equation}
$\mathrm{SU}(2)_{CS}$ contains $\mathrm{U}(1)_A$ as a subgroup.
The $\mathfrak{su}(2)$ algebra
$[\Sigma_\alpha,\Sigma_\beta]=2i\epsilon^{\alpha\beta\gamma}\Sigma_\gamma$
is satisfied with any $k=1,2,3,4$. The  $\mathrm{SU}(2)_{CS}$ transformations mix the
left- and right-handed components of the quark field. It is not a symmetry
of the free massless quark Lagrangian.
For $z$-direction correlators the following representations
of $\mathrm{SU}(2)_{CS}$ are relevant:
\begin{align}
R_1:\;
\{\gamma_1,-i\gamma_5\gamma_1,\gamma_5\}
\\
R_2:\;
\{\gamma_2,-i\gamma_5\gamma_2,\gamma_5\}
\end{align}

These $R_1$ and $R_2$ $\mathrm{SU}(2)_{CS}$ transformations connect the following
operators from the $E_2$ multiplet:
\begin{align}
R_1: &\quad
V_y \leftrightarrow T_t \leftrightarrow X_t, 
\\
R_2: &\quad
V_x \leftrightarrow T_t \leftrightarrow X_t,
\end{align}
as well as  the operators from the $E_3$ multiplet:
\begin{align}
R_1: &\quad
V_t \leftrightarrow T_y \leftrightarrow X_y,
\\
R_2: &\quad
V_t \leftrightarrow T_x \leftrightarrow X_x.
\end{align}

Our symmetry group $D_{4h}$ includes both the permutation operator 
$\hat P_{xy}$ and $\mathds{1}$ transformations, which form a group
$S_2$.
$\hat P_{xy}$ permutes $\gamma_1$ and $\gamma_2$, and transforms
$\gamma_5$ to $-\gamma_5$.
Then $P_{xy} R_1$ is isomorphic to $R_2$. 
This means that $S_2 \times \mathrm{SU}(2)_{CS}$ contains multiplets
\begin{gather}
(V_x,V_y,T_t,X_t), \\
(V_t,T_x,T_y,X_x,X_y).
\end{gather}
The degeneracy between $\mathbf{V}$ and $\mathbf{A}$ means 
$\mathrm{SU}(2)_L \times \mathrm{SU}(2)_R$ symmetry. A minimal group that includes 
$\mathrm{SU}(2)_L \times \mathrm{SU}(2)_R$ and $\mathrm{SU}(2)_{CS}$ is $\mathrm{SU}(4)$.
The $15$ generators of $\mathrm{SU}(4)$ are the following matrices:
\begin{align}
\{
(\tau_a \otimes \mathds{1}_D),
(\mathds{1}_F \otimes \Sigma_i),
(\tau_a \otimes \Sigma_i)
\}
\end{align}
with flavour index $a=1,2,3$ and $\mathrm{SU}(2)_{CS}$ index $i=1,2,3$.
Predictions of $S_2 \times \mathrm{SU}(4)$ symmetry for isovector operators are the
following multiplets:
\begin{align}
(V_x,V_y,T_t,X_t,A_x,A_y), \\
(V_t,T_x,T_y,X_x,X_y,A_t).
\label{eq:su4multiplet}
\end{align}
$S_2 \times \mathrm{SU}(4)$ multiplets include in addition the isoscalar partners of
$V_x$, $V_y$, $T_t$ and $X_t$ operators for the first multiplet in
(\ref{eq:su4multiplet})
as well as of $V_t$, $T_x$, $T_y$, $X_x$,$X_y$ for the second multiplet in
(\ref{eq:su4multiplet}).

\section{\label{sec:conclusions}Conclusions}

In this work we studied the high temperature behaviour of
spatial correlation functions for connected spin $J=0$ and
$J=1$ quark bilinears. Above the critial temperature we see
restoration of both chiral symmetries, $\mathrm{U}(1)_A$ and
$\mathrm{SU}(2)_L \times \mathrm{SU}(2)_R$. Additionally, some unexpected observations are made:
the approximate symmetries
$\mathrm{SU}(2)_{CS}$ and $\mathrm{SU}(4)$ emerge as the temperature is increased
to $\sim 2T_c$. At the same time there are no indications that the
data would converge to the naive free quark limit.

For a discussion of implications of these observations see ref \cite{Rohrhofer:2017grg}. 



\subsection*{Acknowledgments}
We thank C. Gattringer for numerous discussions.
Support from the Austrian Science Fund (FWF) through the grants
DK W1203-N16 and P26627-N27 is acknowledged.
Numerical calculations are performed on Blue Gene/Q at KEK under its 
Large Scale Simulation Program (No. 16/17-14). This work is supported in 
part by JSPS KAKENHI Grant Number JP26247043 and by the Post-K 
supercomputer project through the Joint Institute for Computational 
Fundamental Science (JICFuS).
G.C. is supported by STFC, grant ST/L000458/1.
S.P. acknowledges the financial support from the Slovenian Research
Agency ARRS (research core funding No. P1-0035).

\bibliography{lattice2017}

\begin{thebibliography}{21}

\bibitem{Cossu:2013uua}
G.~Cossu, S.~Aoki, H.~Fukaya, S.~Hashimoto, T.~Kaneko, H.~Matsufuru, J.I.
  Noaki, Phys. Rev. \textbf{D87}, 114514 (2013), [Erratum: Phys.
  Rev.D88,no.1,019901(2013)], \texttt{1304.6145}

\bibitem{Tomiya:2016jwr}
A.~Tomiya, G.~Cossu, S.~Aoki, H.~Fukaya, S.~Hashimoto, T.~Kaneko, J.~Noaki,
  Phys. Rev. \textbf{D96}, 034509 (2017), \texttt{1612.01908}

\bibitem{Bazavov:2012qja}
A.~Bazavov et~al. (HotQCD), Phys. Rev. \textbf{D86}, 094503 (2012),
  \texttt{1205.3535}

\bibitem{Cohen:1996ng}
T.D. Cohen, Phys. Rev. \textbf{D54}, R1867 (1996), \texttt{hep-ph/9601216}

\bibitem{Glozman:2014mka}
L.{\relax Ya}. Glozman, Eur. Phys. J. \textbf{A51}, 27 (2015),
  \texttt{1407.2798}

\bibitem{Glozman:2015qva}
L.{\relax Ya}. Glozman, M.~Pak, Phys. Rev. \textbf{D92}, 016001 (2015),
  \texttt{1504.02323}

\bibitem{Denissenya:2014poa}
M.~Denissenya, L.{\relax Ya}. Glozman, C.B. Lang, Phys. Rev. \textbf{D89},
  077502 (2014), \texttt{1402.1887}

\bibitem{Denissenya:2014ywa}
M.~Denissenya, L.{\relax Ya}. Glozman, C.B. Lang, Phys. Rev. \textbf{D91},
  034505 (2015), \texttt{1410.8751}

\bibitem{Denissenya:2015mqa}
M.~Denissenya, L.{\relax Ya}. Glozman, M.~Pak, Phys. Rev. \textbf{D91}, 114512
  (2015), \texttt{1505.03285}

\bibitem{Denissenya:2015woa}
M.~Denissenya, L.{\relax Ya}. Glozman, M.~Pak, Phys. Rev. \textbf{D92}, 074508
  (2015), [Erratum: Phys. Rev.D92,no.9,099902(2015)], \texttt{1508.01413}

\bibitem{Brower:2005qw}
R.C. Brower, H.~Neff, K.~Orginos, Nucl. Phys. Proc. Suppl. \textbf{153}, 191
  (2006), \texttt{hep-lat/0511031}

\bibitem{Brower:2012vk}
R.C. Brower, H.~Neff, K.~Orginos (2012), \texttt{1206.5214}

\bibitem{Cossu:2015kfa}
G.~Cossu, H.~Fukaya, S.~Hashimoto, A.~Tomiya (JLQCD), Phys. Rev. \textbf{D93},
  034507 (2016), \texttt{1510.07395}

\bibitem{DeTar:1987xb}
C.E. Detar, J.B. Kogut, Phys. Rev. \textbf{D36}, 2828 (1987)

\bibitem{Kogut:1998rh}
J.B. Kogut, J.F. Lagae, D.K. Sinclair, Phys. Rev. \textbf{D58}, 054504 (1998),
  \texttt{hep-lat/9801020}

\bibitem{Karsch:2003jg}
F.~Karsch, E.~Laermann (2003), \texttt{hep-lat/0305025}

\bibitem{Cheng:2010fe}
M.~Cheng et~al., Eur. Phys. J. \textbf{C71}, 1564 (2011), \texttt{1010.1216}

\bibitem{Ishikawa:2017nwl}
K.I. Ishikawa, Y.~Iwasaki, Y.~Nakayama, T.~Yoshie (2017), \texttt{1706.08872}

\bibitem{Gavai:2008yv}
R.V. Gavai, S.~Gupta, R.~Lacaze, Phys. Rev. \textbf{D78}, 014502 (2008),
  \texttt{0803.1368}

\bibitem{Karsch:2000ps}
F.~Karsch, E.~Laermann, A.~Peikert, Phys. Lett. \textbf{B478}, 447 (2000),
  \texttt{hep-lat/0002003}

\bibitem{Rohrhofer:2017grg}
C.~Rohrhofer, Y.~Aoki, G.~Cossu, H.~Fukaya, L.{\relax Ya}. Glozman,
  S.~Hashimoto, C.B. Lang, S.~Prelovsek (2017), \texttt{1707.01881}

\end{thebibliography}

\end{document}